\documentclass{article}
\usepackage[margin=1in]{geometry}
\usepackage{amsmath,amssymb,amsthm}
\usepackage{graphicx}
\usepackage{cite}
\usepackage{hyperref}
\usepackage{microtype}

\title{$K$-Best Enumeration}
\author{David Eppstein\thanks{Computer Science Department, University of California, Irvine; Irvine, CA, USA. This material is based upon work supported by the National Science Foundation under Grant CCF-1228639 and by the Office of Naval Research under Grant No. N00014-08-1-1015. A significantly shorter version of this material appears in the Springer \emph{Encyclopedia of Algorithms}, 2014.}}

\date{ }

\begin{document}
\maketitle

\begin{abstract}
We survey $k$-best enumeration problems and the algorithms for solving them, including in particular the problems of finding the $k$ shortest paths, $k$ smallest spanning trees, and $k$ best matchings in weighted graphs.
\end{abstract}

\section{Introduction}
Researchers in the design and analysis of algorithms have had much success in finding precise mathematical formulations of optimization problems, developing efficient algorithms for solving those problems, and proving worst-case bounds on the time complexity of these algorithms. A case in point is the problem of finding minimum spanning trees in graphs, studied since the 1920s~\cite{Bor-PB3-26} and finally shown in 1995 to be solvable in (randomized) linear time~\cite{KarKleTar-JACM-95}. Other problems of a similar nature are the graph shortest path problem, famously solved by Dijkstra's algorithm~\cite{Dij-NM-59}, and the problem of finding minimum weight matchings in graphs, first solved in polynomial time by Edmonds~\cite{Edm-JNBS-65} and later significantly improved~\cite{GalMicGab-SJC-86,GabTar-JACM-91}.

However, many real-world problems are only approximately modeled by these mathematical formulations. The most desirable solution may depend on quality criteria that are more complicated than a simple sum of edge weights, it may be determined by data that is not yet available or subject to unpredictable changes, or its choice may involve political or social processes that are not well-modeled by mathematical values. In such a situation, it is desirable for an algorithm to output more than one candidate solution. Once such a list of candidates is generated, one can apply more sophisticated quality criteria, wait for data to become available to choose among them, or present them all to human decision-makers. On the other hand, the exponential growth of the solution space for many combinatorial optimization problems would make it infeasible to list all possible candidate solutions; instead, some filtering is needed.

A common approach to such problems breaks the decision process into two stages. In the first stage, the problem is modeled mathematically in the standard way (e.g. for the problems above, by weighting the edges of a graph and seeking a solution with a small sum of edge weights). However, rather than finding a single optimal solution, an algorithm is used to find the \emph{$k$ best solutions}, for a given parameter value $k$:  that is, the algorithm finds a set of $k$ different solutions that have a better solution quality than any solution not in the set. Then, these candidates are passed on to the second stage of the decision process, where they may be evaluated and compared using more complicated evaluation criteria than sums of weights, used as a set of candidates to switch among dynamically based on updated data, or passed to human decision-makers. A familiar example of this occurs in car navigation systems, which can typically be programmed to present drivers with multiple alternative routes between the current location of the car and the target destination.

In this review we survey algorithms for generating sets of the $k$ best solutions, for several optimization problems, as well as the applications of these algorithms.

\section{General Principles}

If the single best solution to an optimization problem can be found by an algorithm whose running time is polynomial in the input size, then it is typical for the $k$ best solutions to be found in time polynomial in both the input size and $k$. There are two general techniques that can be used to achieve this, based on optimal substructures and solution space partitioning. In turn, these methods rely on fast algorithms for the \emph{selection problem} in certain sets of structured values.

\subsection{Selection}
Underlying many $k$-best algorithms is the problem of \emph{selection}, finding the $k$ smallest values from a larger set of values~\cite{BluPraTar-JCSS-73}. If there are $n$ values in the set, then selection may be solved in time $O(n+k)$, for instance by \emph{quickselect}, an algorithm that chooses a random pivot value from the set, partitions the rest of the set into subsets that are less than, equal to, or greater than the pivot, and then recurses into one of these subsets~\cite{MarRou-SJC-01}.

However, in $k$-best problems, we do not wish to generate all solutions before finding the best of them: we wish to avoid the $O(n)$ part of this $O(n+k)$ time bound. Often, the solution values among which we are selecting the $k$ best have some additional structure that allows finding the $k$ best without examining all solutions. For instance, suppose that the space of solutions can be represented as the vertex set of an (implicitly defined) edge-weighted graph, that the degrees of the vertices in this graph are bounded, and that the quality of any particular solution equals the length of the shortest path in this graph from some designated starting vertex to the solution vertex. In such a case, Dijkstra's algorithm can be used to recover the $k$ smallest of these values in time $O(k\log k)$, independent of the size of the graph, without requiring that the whole graph be explored. Other structures that allow the $k$ best solutions to be found more efficiently than enumeration of the whole solution set include representations of the solution space as the set of sums of pairs of values drawn from two sorted sets~\cite{FreJoh-JCSS-82,JohKas-JACM-78}, or as the elements of an (implicitly defined) matrix whose rows and columns are sorted~\cite{FreJoh-SJC-84}. Again, for such structures, the $k$ smallest values can be found efficiently without examining the whole matrix or the whole set of pair sums.

A general framework that can be used to describe many of these structured selection problems is the problem of selection in a $d$-ary heap. A min-heap is a rooted tree, with values associated with its nodes, such that each non-root node has a value that is at least as large as its parent. In order to avoid having to construct all solutions before performing a selection algorithm, and in order to handle situations in which this tree has infinitely many nodes, we will assume that our selection algorithm is given as input an implicit representation of such a heap.
That is, its input consists of a binary representation of the root node of the heap together with pointers to two subroutines that take a single node as input: one subroutine to list the children of any node in the heap, and another subroutine to find the value of any node in the heap. Based on this information, the problem is to find the $k$ nodes with the smallest values. For instance, the problem of selection in a sorted array can be represented in this way by a heap in which each node represents an array cell, the root is the upper left corner of the array, and the parent of each node is either the cell above it (if it is not in the first row of the array) or to its left (if it is in the first row). An implicit min-heap represented in this way is a special case of a graph (where the weight of an edge is the difference in values between a node and its parent), so Dijkstra's algorithm can find the $k$ best solution values in such a structure in time $O(k\log k)$. A more sophisticated algorithm of Frederickson for searching an implicit heap improves this time bound to $O(k)$~\cite{Fre-IC-93}.

\subsection{Optimal Substructures}
The optimal substructure technique is most applicable to problems whose optimization version is solved by a dynamic programming algorithm. In the dynamic programming technique, one identifies a family of polynomially many subproblems of the same type as the original problem, and a recurrence relation by which the value of any subproblem can be calculated in terms of the values of smaller subproblems.  The value of the whole problem can then be found by iterating through all of the subproblems in order by their sizes, using the recurrence to calculate and store the value of each subproblem as a combination of previously-computed values. To apply this algorithmic framework to a $k$-best problem, one stores the $k$ best solutions to each subproblem (rather than a single best solution), and modifies the recurrence to compute these $k$ best solutions as combinations of the $k$ best solutions of smaller subproblems. The $k$-best algorithm then iterates through the subproblems in order by size, using this recurrence to compute the $k$ best values of each subproblem.

As an example, the problem of finding a minimum weight triangulation of a point set (NP-hard for arbitrary planar point sets~\cite{MulRot-JACM-08}) can be solved in polynomial time for points in convex position, by a dynamic program in which the subproblems are the subsets of the points that can be formed by intersecting them with a half-plane~\cite{Gil-MS-79,Kli-ADM-80}. The optimal solution for such a problem may be found by choosing one edge of the convex hull, testing all triangles that could be based on that edge, and for each such triangle adding the weights of the optimal solutions of the two subproblems that the triangle splits the problem into. For $n$ input points, there are $O(n^2)$ subproblems, the solution to each of which involves examining $O(n)$ triangles, so the total time is $O(n^3)$. To modify this dynamic programming algorithm to find the $k$ smallest-weight triangulations, we may store the $k$ smallest weight solutions for each subproblem (in sorted order). To compute the value of a problem, we again choose one convex hull edge and test all triangles that include that edge. However, a single triangle may produce $O(k^2)$ solutions (combinations of the $k$ values stored in each of the two subproblems on either side of the triangle), so implemented naively this method would take $O(k^2 n)$ time per subproblem to sort through all these solutions and select the $k$ best of them, giving an $O(k^2 n^3)$ overall time bound. To speed this up, one can replace the computation for each triangle by an algorithm for finding the $k$ smallest values among the pairwise sums from two sets of $k$ sorted values, which may be solved in $O(k)$ time~\cite{FreJoh-JCSS-82}. This speedup leads to an $O(kn^3)$ time bound overall.

\subsection{Solution-Space Partitions}

An alternative general approach to $k$-best problems involves partitioning the space of solutions into smaller subspaces defined by subproblems of the original problem. For instance, in a problem where the solutions may be described as sets of edges in a graph, a subproblem may be specified by requiring some edges to be included in any solution while removing some other edges from the graph, forcing them to be excluded.

If it is possible to find not just the best solution to a given optimization problem, but also the second best, then this may be used to partition the space of solutions into a collection of subproblems that have the structure of a binary heap. Each node of this heap represents a single subproblem (represented e.g. by its sets of forbidden and required edges) and has a value equal to the second-best solution within that subspace. The root of the heap is the space of all solutions, i.e., a subproblem where we have not yet made any restrictions. For a given node $X$ of this heap, the best and second-best solution must differ from each other, for instance by including an edge $e$ in one of these two solutions and excluding $e$ from the other solution. The two children of node $X$ may then be formed as the subproblems where $e$ is added respectively to the set of forbidden or required edges. One of these two subproblems includes the best but not the second best solution, and the other includes the second best but not the best solution, so both subproblems have second-best solution values that differ from the value for the whole space. This hierarchical partition of the solution space organizes all the solutions except the best one into an implicit binary heap, allowing the selection algorithm of Frederickson~\cite{Fre-IC-93} to be used to find the $k$ best solutions while examining only $O(k)$ of the subproblems from this structure. This, the time for this procedure is $O(k)$ times the time to find a single second-best solution.

An alternative partitioning technique uses only the best solution in each subproblem, rather than also requiring the second best solution. However, in exchange for this easier computation in each subproblem, it produces a heap-ordered tree of solution values in which the degree of each tree node is higher. Suppose, for instance, that the $k$-best problem that we are trying to solve has an input in the form of a graph, and that its solutions can be represented by sequences of edges in this graph; for instance, this is true of the $k$-shortest paths and $k$-best spanning trees problems.
Suppose that the best solution to the problem is represented by the sequence of edges $e_1,e_2,\dots$. Then, from this solution, we form a collection of subproblems, one for each of these edges, where the $i$th subproblem consists of the solutions that are forced to include all the edges in the sequence up to $e_{i-1}$ but that exclude $e_i$. Every solution to the overall problem, other than the best solution, belongs to exactly one of these subproblems: the one defined by the first difference between the given solution and the best solution. Recursively subdividing each of these subproblems into sub-subproblems, etc., produces a heap-ordered tree whose degree equals the maximum number of edges. Again, applying a selection algorithm in a heap allows the $k$ best solutions to be found. However, because of the higher degree of the min-heap in this technique, the number of subproblems that will be examined is $O(ks)$ and the overall time is $O(ks)$ times the time for finding a single best solution, where $s$ is the maximum number of edges in a single solution.

These two partitioning techniques are reviewed in more detail by Hamacher and Queyranne~\cite{HamQue-AOR-85}. They attribute the binary heap partition method to a $k$-best spanning tree algorithm of Gabow~\cite{Gab-SJC-77}. The multiway partition based only on the best solution comes from a $k$-best matching algorithm by Murty~\cite{Mur-OR-68} and its generalization by Lawler~\cite{Law-MS-72}.

\section{Shortest Paths}
Probably the most important and heavily studied of the $k$-best optimization problems is the problem of finding $k$ shortest paths~\cite{HofPav-JACM-59,BelKal-JSIAM-60,Pol-JMAA-61,Pol-OR-61,Pol-ORSA-69,Law-MS-72,Shi-JNBS-74,Shi-Nw-76,Won-Nw-76,Law-CACM-77,Shi-Nw-79,CarSod-MOR-83,Fuc-ORP-85,SkiGol-Nw-87,LalVas-GVUZPM-87,LawRez-MCNMC-93,ChoMadMor-JCI-95,Epp-SJC-98,GueMus-JOTA-00,Rup-Algo-00,RinRodSun-AML-00,GueMusLacPec-OR-01,JimMar-WEA-03,CheYan-COR-04,AndZog-ORL-08,GraChe-JGT-11,AljLeu-AI-11,LiuRam-INFOCOM-01,JimMar-WAE-99,AzeSanSil-EJOR-93,vdZFio-TRB-05,MarPasSan-IJFCS-99,BraSin-UKPEW-96,Che-COR-93,Aze-EJOR-94,MinShi-JIMA-73,Min-CACM-74,Min-JNM-75,GunSchSie-DYADEM-10,Bei-Comp-72,Fox-CACM-75,Fox-INFOR-73,Fox-BORSA-75,JinCha-ISCS-89,Kun-Int-94,MahHuZil-CSOR-92,Mar-IEICE-93,SkiGol-AOR-89,Wei-Comp-76,Wei-Nw-73}.

The $k$-shortest paths problem includes as special cases finding the $k$ best solutions to problems such as biological sequence alignment~\cite{ShiIma-PSB-97,ShiIma-JCB-97,ByeWat-OR-84,NaoBru-JCB-94,Wat-PNAS-83} or the $(0,1)$-knapsack problem~\cite{YanSomMac-PO-00}, whose dynamic programming solutions can be expressed as shortest path problems in an associated graph.
Many problems of hypothesis generation in natural language processing and speech recognition can also be formulated as $k$-best optimization problems~\cite{Fil-COLING-10,KniGra-CL-98,KniHat-ACL-95,BetHil-ASSP-95,CheSoo-TSAP-94,ChoLeeJua-ICASSP-93,ChoLeeJua-PRAI-94,ChoMatJua-ASSP-94,ChoSch-SNLW-89,SchCho-ICASSP-90,SchAus-ICASSP-91,DaiLee-ICCC-94,HisNit-SCJ-95,ChaJoh-ACL-05,SooHua-ICASSP-91}.
The Viterbi decoding technique for Markov models, commonly used to model these problems, can also be formulated as a search for a path in an associated graph, with a vertex for each pair of a time step and a Markov state, and the $k$-best beam search technique used for multiple hypothesis generation in these problems can  be interpreted as a special case of a $k$-shortest-path algorithm. This method can also be used to combine different techniques for language and speech recognition, by using one technique to generate hypotheses and the other technique to rescore them~\cite{OstKanAus-HLT-91}.

The many other applications of the $k$ shortest paths problem include reconstruction of metabolic pathways~\cite{Ari-SPT-00} and gene regulation networks~\cite{ShiPar-BI-12},
motion tracking~\cite{BerFleTur-PAMI-11},
message routing in communications networks~\cite{BelKal-JSIAM-60,Top-TC-88,MaoZhaHua-JWU-13,BalSteBal-INFOCOM-91,BalSteSim-TN-95},
listing close genealogical relationships in highly intermarried pedigrees~\cite{Epp-SJC-98},
power line placement~\cite{ChoBur-TPAS-84},
vehicle and transportation routing~\cite{HadChrMin-AOR-95,XuHeSon-COR-12,JinCheJia-COR-13,MiaChi-EJOR-91},
building evacuation planning~\cite{KarSmi-EO-84},
timing analysis of circuits~\cite{YenDuGha-DAC-89,AsaSat-GTA-85},
task scheduling~\cite{Hor-JORS-80,Dod-OR-84},
and communications and transportation network design~\cite{DunGroMac-SAC-94,MacGro-SPE-94,BruGhiImp-EJOR-98,BusLocOls-GC-94,DioCliNor-IEE-89},
as well as in subroutines for other combinatorial optimization problems~\cite{ChaLai-JIOS-98,JiaVar-TAC-06,FuRil-TRB-98,BesKel-AWOCA-99,CouCliCur-COR-99,CurReVCoh-TS-87}

In the most basic version of the problem, the input is a weighted directed graph, with two designated source and destination vertices $s$ and $t$, and a number $k$. The goal is to find $k$ different walks (paths allowing repeated vertices) from $s$ to $t$, with the minimum possible weights. An algorithm by Eppstein~\cite{Epp-SJC-98} achieves optimal asymptotic time complexity: $O(m+n\log n+k)$, constant time per path after a preprocessing stage with the running time of Dijkstra's algorithm for a single shortest path. Eppstein's algorithm begins by computing a tree $T$ of shortest paths to $t$, and (following Hoffman and Pavley~\cite{HofPav-JACM-59}) it represents each of its output paths by the sequences of \emph{detours} that these paths make: edges that do not belong to~$T$. The length of the path is then the shortest path distance from $s$ to $t$, plus the sum of the lengths added by each detour. For each vertex $v$ in the graph, Eppstein's algorithm constructs a collection of the detours whose starting vertex is on the  path in~$T$ from $v$ to $t$; this collection is represented as a binary heap, using persistent data structure techniques~\cite{DriSarSle-JCSS-89} to allow these collections to share substructures with each other to save preprocessing time and storage. The algorithm uses these collections to partition the space of solutions into a collection of subproblems that themselves have the structure of a bounded-degree heap. Each subproblem consists of the paths that start with a given sequence of detours, and that use at least one more detour from a given binary heap of detours. The optimal solution of such a subproblem is the one whose final detour is at the root of the heap, and it has three subproblems with worse solutions as children: two in which the root detour is not used and instead the path uses at least one detour from a child of the root in the binary heap of detours, and one where the root detour is used but is not the last detour, and the next detour comes from the binary heap associated with the endpoint of the root detour.

In a graph with cycles, the $k$ shortest paths may consist of as few as one simple path, together with one or more repetitions of a short cycle starting and ending at one of the vertices of the path. Loops of this type are generally not desired, and the problem is particularly critical when the input is an undirected graph, as converting it to a directed graph by replacing each undirected edge by two directed edge will create many potential loops. Beginning with Clarke, Krikorian, and Rausen~\cite{ClaKriRau-JSIAM-63} researchers have developed algorithms that instead seek the $k$ shortest simple (or loopless) paths from $s$ to $t$ in a network~\cite{Yen-MS-71,Per-Nw-86,HerMaxSur-TAlg-07,Rod-SJC-10,Che-IPL-94,MarPas-4OR-03,Ber-SODA-10,CarWoo-Nw-05,Ish-JORSJ-78,SugKat-IPSJ-85}. Yen's algorithm~\cite{Yen-MS-71} still remains the one with the best asymptotic time performance. It is based on best-solution partitioning and Dijkstra's algorithm; the number of edges in a single solution is at most $n-1$, and the time to find a solution using the Fibonacci-heap variant of Dijkstra's algorithm is $O(m+n\log n)$ (in a graph with $m$ edges and $n$ vertices), so following the general form for best-solution partitioning, the time for this method is $O(kn(m+n\log n)$. A more recent algorithm of Hershberger, Maxal, and Suri~\cite{HerMaxSur-TAlg-07} is often faster, but is based on a heuristic that can sometimes fail, causing it to fall back to Yen's algorithm and achieve the same performance bound. In the  case of undirected graphs, it is possible to find the $k$ shortest simple paths faster, in time $O(k(m+n\log n))$~\cite{KatIbaMin-ECJ-78,KatIbaMin-Nw-82,HadChr-Nw-99}.

Minieka~\cite{Min-CACM-74} and Fox~\cite{Fox-CACM-75} considered an all-pairs variant of the $k$-shortest-paths problem in which the goal is to find a separate set of $k$ paths for each pair of vertices in the graph. For this problem, Eppstein's algorithm requires only $n$ copies of the preprocessing stage (one for each destination vertex), after which each path takes constant time to find, so the total time is $O(mn+n^2\log n+kn^2)$.
The shortest path tree in a graph is a tree connecting a given source vertex to all other vertices, minimizing the sum of the path costs to the other vertices. The $k$-best version of this problem, seeking the $k$ best trees according to this quality measure, has also been studied~\cite{SedGon-EJOR-10}.

There has also been research on finding a given number of paths between two given terminals that are completely disjoint from each other and minimize a sum of weights~\cite{Cas-TAES-90,NikPitTip-INFOCOM-97}. This problem can be solved as a special case of the minimum-cost flow problem; it has a significantly different flavor from $k$-best enumeration problems, as the choice of one path affects the others and the number of paths that can be selected is much smaller.

\section{Spanning Trees}
A 1977 paper of Gabow~\cite{Gab-SJC-77} introduced both the problem of finding the $k$ minimum-weight spanning trees of an edge-weighted graph, and the technique of finding a binary hierarchical subdivision of the space of solutions, which he used to solve the problem. In any graph, the best and second-best spanning trees differ only by one edge swap (the removal of one edge from a tree and its replacement by a different edge that reconnects the two subtrees formed by the removal), a property that simplifies the search for a second-best tree as needed for Gabow's partitioning technique. For similar reasons, when $k$ is smaller than the numbers $n$ and $m$ of vertices or edges in the input graph, the graph may be simplified by finding a single minimum spanning tree, computing the best swap that each edge of the graph participates in, removing the edges that are not in the tree and do not participate in the $k$ best swaps, and contracting the edges that are in the tree but do not participate in the $k$ best swaps. For this reason, factors of $n$ and $m$ in the running time of any $k$-best spanning tree algorithm may be replaced by $k$ when this replacement would be an improvement~\cite{Epp-BIT-92}.

The fastest known algorithms for the $k$ best spanning trees problem are based on Gabow's partitioning technique, together with dynamic graph data structures that keep track of the best swap in a network as that network undergoes a sequence of edge insertion and deletion operations~\cite{Fre-SJC-85,Fre-SJC-97,EppGalIta-JACM-97}. To use this technique, one initializes a fully-persistent best-swap data structure (one in which each update creates a new version of the structure without modifying the existing versions, and in which updates may be applied to any version~\cite{DriSarSle-JCSS-89}) and associates its initial version with the root of the subproblem tree. Then, whenever an algorithm for selecting the $k$ best nodes of the subproblem tree generates a new node (a subproblem formed by including or excluding an edge from the allowed solutions) the parent node's version of the data structure is updated (by either increasing or decreasing the weight of the edge to force it to be included or excluded in all solutions) and the updated version of the data structure is associated with the child node. In this way, the data structure can be used to quickly find the second-best solution for each of the subproblems explored by the algorithm. Based on this method, the $k$ best spanning trees of a graph with $n$ vertices and $m$ edges can be found (in an implicit representation based on sequences of swaps rather than explicitly listing all edges in each tree) in time $O(\operatorname{MST}(m,n) + k\min(n,k)^{1/2})$ where $\operatorname{MST}(m,n)$ denotes the time for finding a single minimum spanning tree~\cite{EppGalIta-JACM-97}. If randomized algorithms are considered, the minimum spanning tree problem can be solved in linear time~\cite{KarKleTar-JACM-95}, so the $\operatorname{MST}(m,n)$ term can be replaced by $m+n$.
This technique may also be used to find the $k$ best spanning trees of a set of $n$ points in the Euclidean plane, in time
$O(n\log n\log k + k\min(n,k)^{1/2})$~\cite{Epp-IJCGA-94,EppGalIta-JACM-97}; these trees may include pairs of edges that cross each other, but it is also possible to find the $k$ best non-crossing planar spanning trees efficiently~\cite{MarNie-CGTA}.

The $k$th smallest distinct weight of a spanning tree may be obtained by a sequence of at most $k-1$ swaps from any minimum spanning tree, allowing these weights to be generated in polynomial time when $k$ is constant~\cite{Kan-Comb-87,MayPla-Comb-92}. However, when $k$ is an input variable, finding the $k$th smallest distinct spanning tree weight remains $\mathsf{NP}$-hard~\cite{MayPla-Comb-92}.

The problem of finding the $k$ best spanning trees has been applied to $\mathsf{NP}$-hard multicriterion optimization problems for spanning trees~\cite{HamRuh-AOR-94,CliCapPas-EJOR-10,SteRad-COR-08}, to point process intensity estimation~\cite{HerMic-SIT-97}, to the analysis of metabolic pathways~\cite{AriAsaNis-RECOMB-00}, to image segmentation~\cite{StrPetKot-PR-13} and classification~\cite{ElFDaoAbo-IVC-08}, to the reconstruction of pedigrees from genetic data~\cite{Cow-TPB-13},
to the parsing of natural-language text~\cite{Agi-PC2-12}, and to the analysis of electronic circuits~\cite{YuSec-IJCTA-96}.
This problem is a special case of finding the $k$ best bases of a matroid, which has also been studied~\cite{BurHaf-SEC-74,HamQue-AOR-85,LecRen-ZOR-90,Cha-ORL-08}.
For additional research on the $k$ smallest spanning trees problem see \cite{BurHaf-SEC-74,KatIbaMin-SJC-81,MaIwaGu-AIZU-97,TanLia-JCUST-90,BerFilLeo-ICARIS-07,SorJan-PO-05}.

\section{Other Problems}

After paths and spanning trees, probably the next most commonly studied $k$-best enumeration problem concerns matchings.
The problem of finding the $k$ minimum-weight perfect matchings in an edge-weighted graph was introduced by Murty~\cite{Mur-OR-68}. A later algorithm by Chegireddy and Hamacher~\cite{CheHam-DAM-87} solves the problem in time $O(kn^3)$ (where $n$ is the number of vertices in the graph) using the technique of building a binary partition of the solution space.
The $k$-best matchings have been used to find matchings with additional side constraints~\cite{BalDerHil-Nw-90} or with multivariate optimization criteria~\cite{Sch-CA-95}.
They have also been applied for message routing in parallel computing systems~\cite{ChiChe-CSSE-97}.
For additional work on this problem see~\cite{DerMet-CO-92,MatTamIke-DAM-94,PasCapCli-4OR-03,PedNieAnd-COR-08}.

In natural language processing applications, an important generalization of the $k$ shortest paths problem involves finding the $k$ best parse trees of a context-free grammar~\cite{HuaChi-IWPT-05,JimMar-APR-00,ChaJoh-ACL-05,ZhaOep-Car-IWPT-07,PauKle-ACL-09,PauKleQui-ACL-10}.
Other problems whose $k$-best solutions have been studied include
the Chinese postman problem~\cite{SarMat-SJO-93},
the traveling salesman problem~\cite{cdPLibSie-COR-99},
the $k$ best spanning arborescences in a directed network~\cite{Cam-Nw-80},
the matroid intersection problem~\cite{CamHam-SJDM-89,YuSec-TCAD-97},
binary search trees and Huffman coding~\cite{AniHas-SJC-89},
chess strategies~\cite{Alt-ICCA-97},
the $k$ best integer flows~\cite{Ham-AOR-95,HamHus-BAAI-94,SedEsp-COR-13},
the $k$ smallest cuts in a network~\cite{Ham-ORL-82,HamPicQue-ORL-84,HamQue-AOR-85},
and, in probabilistic reasoning, the $k$ best solutions to a graphical model~\cite{Nil-SC-98,FroGlo-NIPS-09,FleRolDec-GKR-11,DecFleMar-AAAI-12}.

For many $\mathsf{NP}$-hard optimization problems, where even finding a single best solution is difficult, an approach that has proven very successful is \emph{parameterized complexity}, in which one finds an integer parameter describing the input instance or its solution that is often much smaller than the input size, and designs algorithms whose running time is a fixed polynomial of the input size multiplied by a non-polynomial function of the parameter value. Chen et al.~\cite{CheKanMen-TCS-13} extend this paradigm to $k$-best problems, showing that, for instance, many $\mathsf{NP}$-hard $k$-best problems can be solved in polynomial time per solution for graphs of bounded treewidth.

\bibliographystyle{abuser}
\bibliography{kbest}

\end{document}